\documentclass[conference,a4paper]{IEEEtran}
\IEEEoverridecommandlockouts
\usepackage{cite}
\usepackage{amsmath,amssymb,amsfonts}
\usepackage{algorithmic}
\usepackage{graphicx}
\usepackage{textcomp}
\usepackage{xcolor}
\usepackage{threeparttable}
\usepackage{multirow}
\usepackage{colortbl}
\usepackage{tabularray}
\usepackage{fancyhdr}

\def\BibTeX{{\rm B\kern-.05em{\sc i\kern-.025em b}\kern-.08em
    T\kern-.1667em\lower.7ex\hbox{E}\kern-.125emX}}

\definecolor{Orange}{rgb}{1.0, 0.5, 0.0}
\definecolor{Purple}{rgb}{0.5, 0.0, 0.5}

\newcommand{\blue}[1]{{\color{black}#1}}
\newcommand{\orange}[1]{{\color{black}#1}}
\newcommand{\purple}[1]{{\color{black}#1}}

\begin{document}

\title{Distributed Delay-Based BIST for Mixed-Signal Circuits in Flexible Electronics\\
}

\fancypagestyle{firstpage}{%
    \fancyhf{}  
    \renewcommand{\headrulewidth}{0pt}  
    \renewcommand{\footrulewidth}{0pt}
    \fancyhead[C]{Accepted for publication at the 31st IEEE European Test Symposium (ETS), May 25-29, 2026.}
}

\newif\ifanonym
\IEEEaftertitletext{\vspace{-1.2\baselineskip}} 

\ifanonym
\author{}
\else
\author{
    \IEEEauthorblockN{
        Paula Carolina Lozano Duarte\IEEEauthorrefmark{1},
        Sule Ozev\IEEEauthorrefmark{2},
        Mehdi Tahoori\IEEEauthorrefmark{1}
    }
    \IEEEauthorblockA{
        \IEEEauthorrefmark{1}Karlsruhe Institute of Technology, DE
        \IEEEauthorrefmark{2}Arizona State University, US
    }
    \IEEEauthorblockA{
        \IEEEauthorrefmark{1}\{paula.duarte, mehdi.tahoori\}@kit.edu
        \IEEEauthorrefmark{2}Sule.Ozev@asu.edu
    }
}

\fi

\bstctlcite{IEEEexample:BSTcontrol}

\maketitle
\begin{abstract}
Flexible electronics (FE) based on indium gallium zinc oxide thin-film transistors
(IGZO-TFTs) are emerging for ultra-low-power wearable applications. However, lack
of packaging, limited pins, and high device variability make conventional Automatic
Test Equipment (ATE) impractical for testing analog/mixed-signal circuits in FE.
This work presents a dual-purpose ring oscillator (RO) and voltage-controlled
oscillator (VCO) serving as functional timing blocks and core structures for a
distributed delay-based BIST framework. The oscillators achieve 1100$\times$ area
reduction and 5600$\times$ lower power than previous IGZO-TFT designs. Lightweight
digital BIST embedded within each RO stage enables stage-wise delay monitoring for
\blue{defect} detection. The BIST achieves 93\% \blue{defect} coverage for
individual \blue{defects} and 88\% for multiple simultaneous \blue{defects}, with
only 3\% power overhead and no external test equipment.
\end{abstract}

\section{Introduction}\label{sec:intro}

Flexible electronics (FE)
have enabled mechanically conformable, lightweight, and low-cost
manufacturing~\cite{EuropracticeFlexibleElectronics,lozano:AKAN}.
Their compatibility with plastic or stretchable substrates allows direct
integration on non-traditional surfaces, making them attractive for wearable
health monitoring~\cite{afentaki2025stress}, large-area sensing
platforms~\cite{Zhang2025},
flexible displays~\cite{Lim2025} and human–machine
interfaces~\cite{Heng:AM2022:FlexHumanMachInterfaces}.

Despite these advantages, FE circuits suffer from substantial device variability
due to low carrier mobility, process fluctuations, and the lack of complementary
devices. This variability directly impacts timing precision, frequency generation,
and analog performance, raising serious challenges for reliable mixed-signal
designs. Furthermore, the absence of \purple{rigid} packaging in FE devices and
the limited number of pins make the use of conventional Automatic Test Equipment
(ATE) both costly and impractical, highlighting the need for integrated,
resource-efficient built-in self-test (BIST) solutions for \purple{mixed-signal}
FE circuits.

\orange{Delay-based testing using ring oscillators (ROs) offers a practical path
toward BIST \purple{for mixed-signal circuits in} FE: ROs produce time-domain
signatures that are highly sensitive to device and interconnect variations.
Monitoring these delay variations allows \blue{detection of defects} in the
oscillator stages or in analog\purple{/mixed-signal} circuits connected to them,
without relying on external \purple{test and measurement} equipment.
Although RO-based BIST has been explored in silicon \purple{VLSI}
technologies~\cite{lu2025testingfaulttolerancetechniques, LiROtesting}, these
approaches do not directly extend to FE, where power budgets are significantly
tighter, and circuit topologies are constrained to unipolar devices.
In this context, frequency-generation blocks such as voltage-controlled oscillators
(VCOs) become attractive platforms for unifying timing generation and test
functionality. Beyond providing clocking or frequency
modulation~\cite{VCO_Bongjun,VCO_igzo_tejaswini}, they can act as embedded
monitors that continuously sense delay variations in both the RO and upstream
analog/mixed-signal circuits.}

\blue{In this work, we extend the VCO-based paradigm to FE and propose a tunable
RO optimized for ultra-low-voltage (0.9V) operation, alongside a distributed delay
BIST integrated within each RO stage. Unlike prior work measuring overall
frequency~\cite{lu2025testingfaulttolerancetechniques, LiROtesting}, our per-stage
monitoring approach enables (1)~\blue{detection of local defects} within the
oscillator, (2)~reuse as a modular \blue{defect} detection unit for upstream
analog/mixed-signal circuits, and (3)~stage-level \blue{defect} isolation without
modifying circuits under test.}

\blue{Deviations in delay serve as indirect indicators of circuit malfunctions,
providing lightweight and effective \blue{defect} sensing. \orange{while adding
only 3\% power overhead and operating with a self-generated clock signal.}
Overall, the proposed architecture achieves frequency tunability, stable operation, ultra-low power consumption, and modular \blue{defect} detection \orange{suitable for distributed deployment across FE systems.}}
\section{Background}
\subsection{Flexible Electronics}\label{sec:FE}

FE refer to systems fabricated on deformable substrates that can bend, stretch,
or twist without loss of function\blue{, making them ideal for wearable devices, physiological monitoring, and human–machine interfaces~\cite{afentaki2025stress,Heng:AM2022:FlexHumanMachInterfaces}.}
FE relies on modifying standard semiconductor processes to accommodate substrates
like polyimide. A key material is IGZO TFT for its high mobility, low-temperature
compatibility, and transparency. However, IGZO supports only n-type conduction,
imposing significant design constraints and often leading to higher static power
consumption~\cite{Lozano:aspdac25:BinCoDesign}.
\blue{From a test perspective, silicon VLSI benefits from sophisticated BIST and DFT infrastructures, whereas FE requires compact, low-speed solutions designed for simplicity and reliable in-field verification.}

\subsection{Ring Oscillators and VCOs in FE}
Several RO and VCO implementations using IGZO TFTs have been
reported~\cite{fastiIC_arun, high-speed_yuanfeng, A-igzo_di}. \blue{While
demonstrating oscillator feasibility, these designs typically require high supply
voltages ($\geq$20V) and offer only fixed frequencies, limiting their use in
adaptive, low-power systems.}
\cite{VCO_igzo_tejaswini} proposed a VCO combining IGZO TFTs with an operational
amplifier and comparator for the oscillation, improving tunability
at the cost of higher complexity. \cite{VCO_Bongjun} presented a double-gate
printed SWCNT oscillator for low-voltage wide-range tuning, though it relies on
hybrid materials. \blue{Despite these advances, existing designs remain constrained
by rigid component requirements or limited tunability at low voltages.}
\blue{The versatility of ROs and VCOs extends beyond timing generation: their
inherent delay characteristics can be leveraged for in-situ \blue{defect}
detection, making them natural candidates for BIST architectures.}

\subsection{Built-In Self-Test Architectures}
In mixed-signal circuits, traditional test approaches rely on parametric
measurements compared to designer-specified limits~\cite{bilgic2022performance},
often incurring high development costs and requiring expensive
equipment~\cite{Oshita2016}. BIST techniques reduce reliance on external testers
and enable increased coverage with minimal overhead~\cite{Jeong2016}, though
conventional silicon BIST implementations often require resource-intensive ADCs
and are designed for specific functional blocks, limiting reuse.

Recent studies have explored RO-based BIST
schemes:~\cite{lu2025testingfaulttolerancetechniques} proposes an RO-based
\blue{defect} testing approach for MWCNT interconnects (45.5–100\%
coverage); and~\cite{LiROtesting} introduces an oscillation-ring test scheme for
silicon SoCs (100\% coverage).
\blue{These works confirm the versatility of RO-based test but remain focused on interconnect-level reliability or silicon platforms.}
\section{RO-Based VCO and Distributed Delay BIST}\label{sec:metho}

To address FE test limitations, we propose a fully digital, RO-based BIST with
distributed delay monitoring. Unlike approaches measuring only overall frequency,
our BIST embeds lightweight logic at each RO stage to capture per-stage delays,
enabling: \textbf{(1)~Self-test:} detecting \blue{defects} within the oscillator,
and \textbf{(2)~Upstream monitoring:} reusing the VCO+BIST for analog/mixed-signal
circuits. 

\begin{figure}[h!]
    \centering
    \includegraphics[width=0.95\linewidth]{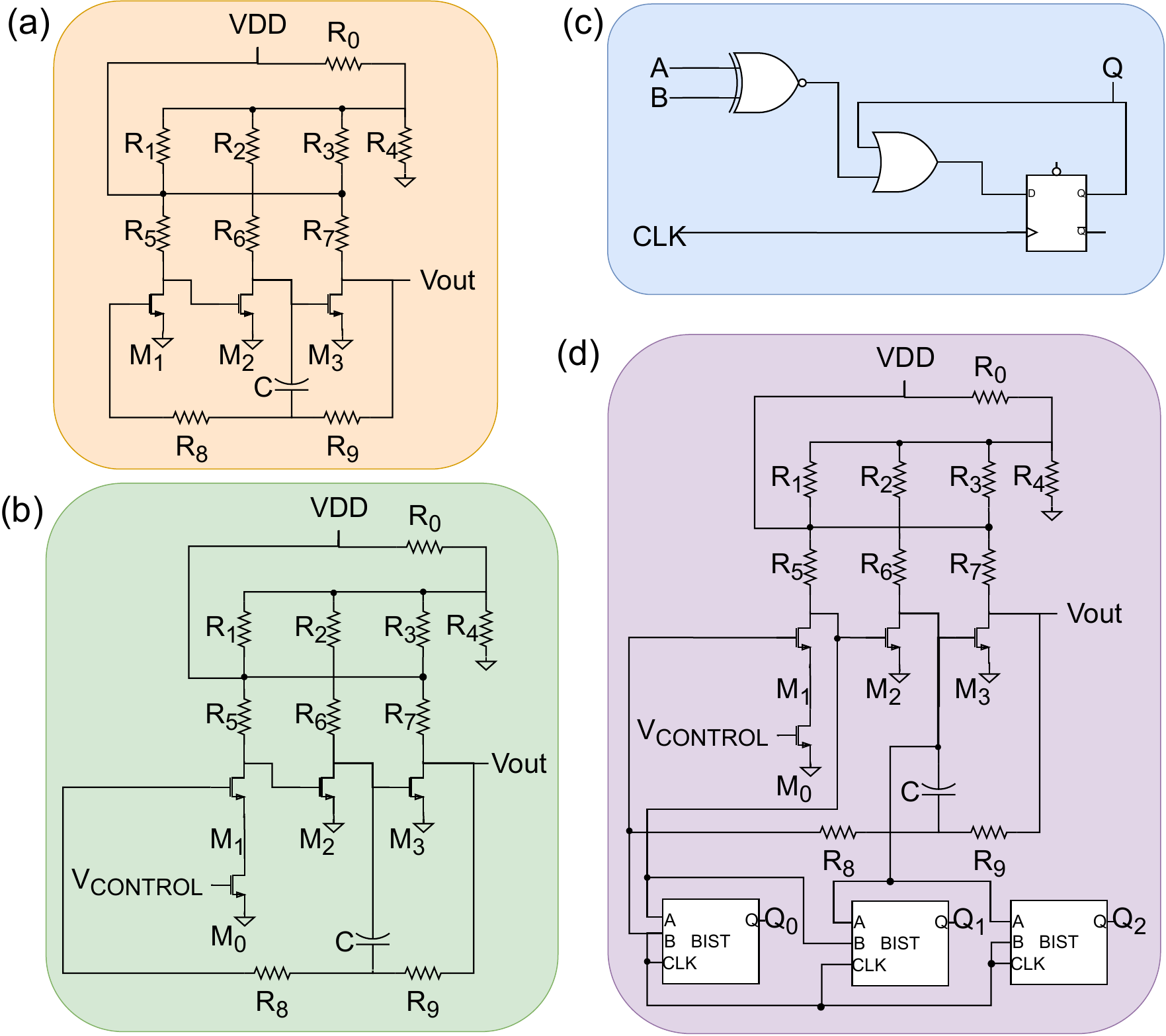}\vspace{-2ex}
    \caption{Schematics of (a)~RO, (b)~VCO, (c)~BIST block, and (d)~integrated
    VCO+BIST. \blue{Logic gates in (c) use standard cells from the FlexIC PDK.}}
    \label{fig:schematic}
    \vspace{-3ex}
\end{figure}

\noindent\textit{RO:} Uses only n-type IGZO TFTs with three cascaded inverting stages ($M_1$–$M_3$).
Pull-up functionality is realized through resistive loads ($R_5$–$R_7$). Series
resistors ($R_1$–$R_3$) are biased from a voltage divider ($R_0$, $R_4$),
stabilizing operation and ensuring sufficient gain. Stages 1 and 3 connect
pull-ups to $V_{DD}$; stage 2 operates from the divided node, introducing
asymmetry that avoids DC locking. Resistors $R_8$, $R_9$ and capacitance $C$
determine oscillation frequency.

\noindent\textit{VCO:}
The VCO introduces a control terminal ($V_{\text{CONTROL}}$) in stage 1. The
source of $M_1$ connects through an additional transistor whose gate is driven by
$V_{\text{CONTROL}}$, modulating the discharge path. Lower $V_{\text{CONTROL}}$
values yield higher frequencies; higher values slow the response. This achieves
wide voltage-controlled tunability with one extra transistor.

\textit{Digital BIST:}
The BIST employs lightweight digital logic to detect \blue{defects} without
external equipment. The test principle compares consecutive stage delays:
consistent phase shifts indicate correct operation; deviations indicate
\blue{defects} in the oscillator or upstream analog/mixed-signal circuits.

Adjacent stage outputs are processed via XNOR gates, followed by an OR gate and
flip-flop for synchronization. The flip-flop uses the oscillator's own output as
clock, eliminating external clock requirements. \blue{Under nominal operation the
BIST output is logic~0 (no \blue{defect} detected). Loss of oscillation—e.g., due
to a catastrophic \blue{defect}—results in a static output treated as a
\blue{defect} condition by default.}

\textit{Full Integration:}
The architecture integrates VCO and BIST (Fig.~\ref{fig:schematic}d). BIST
circuits at each stage enable local delay monitoring with stage-level \blue{defect}
resolution. Stage-wise integration minimizes external I/O and leverages shared
logic for low power and small area.

\textit{\blue{Defect} Injection Strategy.}
To evaluate \blue{defect} tolerance, we target resistive pull-ups in the RO
inverters. \blue{Defects} are injected independently at each resistor for
stage-level observability. Two \blue{defect} types: short circuit (SC) and open
circuit (OC). For SC, nominal resistance (172\,k$\Omega$) is replaced with
10\,k$\Omega$\blue{—modeling resistive bridging while remaining within realistic
process-induced shorts~\cite{spence:analogfaultsimulation}.} For OC, resistance
increases to 10\,M$\Omega$, modeling a broken interconnect. \blue{Opens in long
pull-up resistors are more likely given thin-film resistor geometry; SC
\blue{defects} model resistive bridging during fabrication.} \blue{Defects} are
injected one at a time and in multiple simultaneous combinations.
\section{Simulation Results and Evaluation}\label{sec:results}

\begin{table}
\vspace{-2ex}
\caption{Device properties used for the RO, VCO, and BIST circuits}
\label{tab:sizing}
\scalebox{1}{\resizebox{\linewidth}{!}{%
\centering
\setlength{\arrayrulewidth}{0.4pt} 
\begin{tblr}{
  colspec = {Q[25] Q[25] Q[25]}, 
  vline{2,3} = {-}{0.4pt},
  hline{1,2,15} = {1.2pt}, 
  rowsep = 1.1pt, 
  hline{8,14} = {-}{0.4pt}, 
  hline{3,4,5,6,7,9,10,11,12,13} = {2,3}{0.4pt}, 
}
\textbf{Component} & \textbf{Element} & \SetCell[c=1]{c} \textbf{Size} \\ 

\textbf{RO} & R0 & r = 50$k\Omega$\\
 & R1:R7 & r = 172.47$k\Omega$\\
 & R8 & r = 120$k\Omega$\\
 & R9 & r = 5.6$M\Omega$\\
 & M1:M2 & W= 20$\mu$m L= 600nm \\
 & C & cap = 1pF \\
\textbf{VCO} & R0 & r = 50$k\Omega$\\
 & R1:R7 & r = 172.47$k\Omega$\\
 & R8 & r = 120$k\Omega$\\
 & R9 & r = 5.6$M\Omega$\\
 & M0:M2 & W= 20$\mu$m L= 600nm \\
 & C & cap = 2pF \\
\textbf{BIST} & \SetCell[c=2]{c} Std-cell based design with Gen-3 FlexIC\\
\end{tblr}
}}
\vspace{-2ex}
\end{table}

\begin{figure}
    \centering \includegraphics[width=1\linewidth]{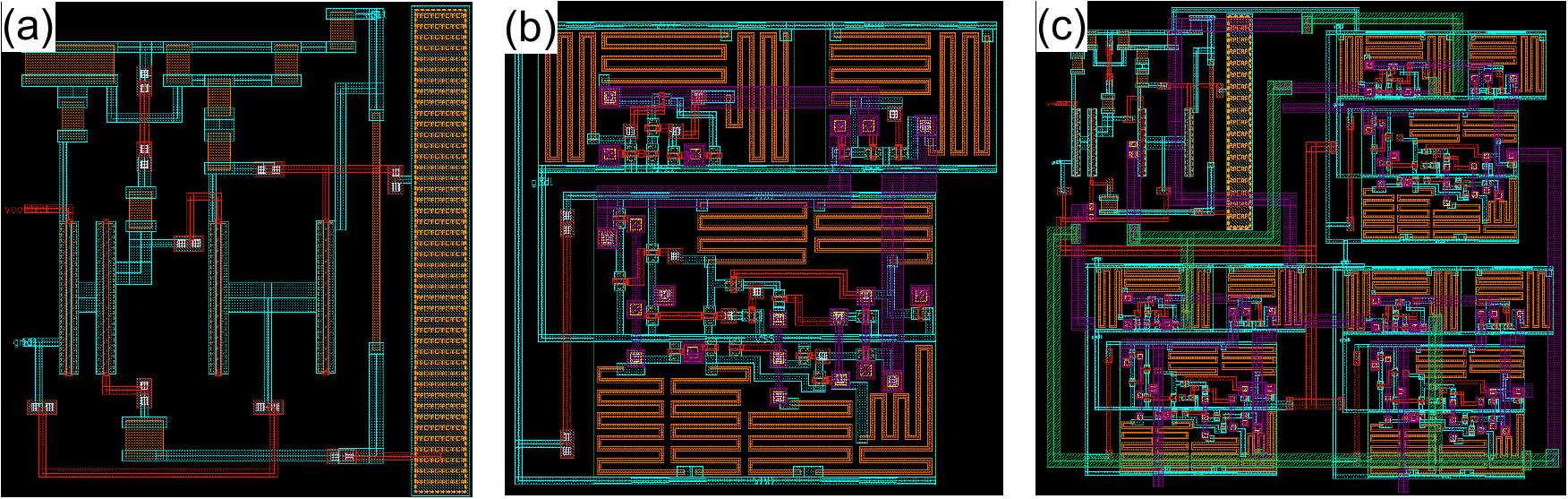}\vspace{-1ex}
    \caption{Layout views of (a) VCO, (b) BIST, and (c) the integrated VCO–BIST design.}
    \label{fig:layouts}
    \vspace{-3ex}
\end{figure}

\begin{table}
\vspace{-1ex}
\caption{Results of the PVT corner analysis simulations}
\label{tab:pvt}
\scalebox{1}{\resizebox{\linewidth}{!}{%
\centering
\setlength{\arrayrulewidth}{0.4pt} 
\begin{tblr}{
  colspec = {Q[100] Q[125] Q[100]}, 
  vline{2} = {2,3,4}{0.4pt}, 
  hline{1,2,5} = {1.2pt}, 
  rowsep = 1.1pt, 
  hline{3,4} = {-}{0.4pt}, 
}
\textbf{Transistor Process} & \textbf{Frequency Range (kHz)} & \textbf{Power Range ($\mu$W)} \\
\textbf{Fast (F)} & 38–122 & 7.1 – 32.8 \\
\textbf{Typical (T)} & 37–97 & 5.7 – 23.9 \\
\textbf{Slow (S)} & 37–85 & 5 – 17.5 \\
\end{tblr}
}}
\vspace{-3ex}
\end{table}

\begin{figure}[t]
    \centering \includegraphics[width=1\linewidth]{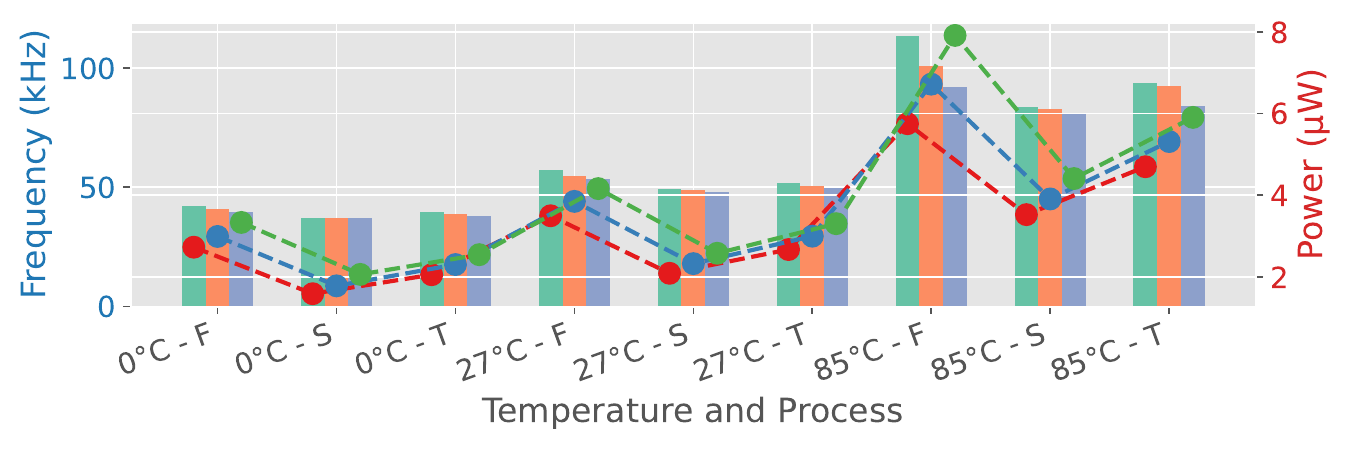}\vspace{-2ex}
    \caption{Effect of $V_{\text{DD}}$ on Frequency and Power per Process-Temperature Corner.}
    \label{fig:PVT_figure}
    \vspace{-3ex}
\end{figure}


\begin{figure}
    \centering \includegraphics[width=1\linewidth]{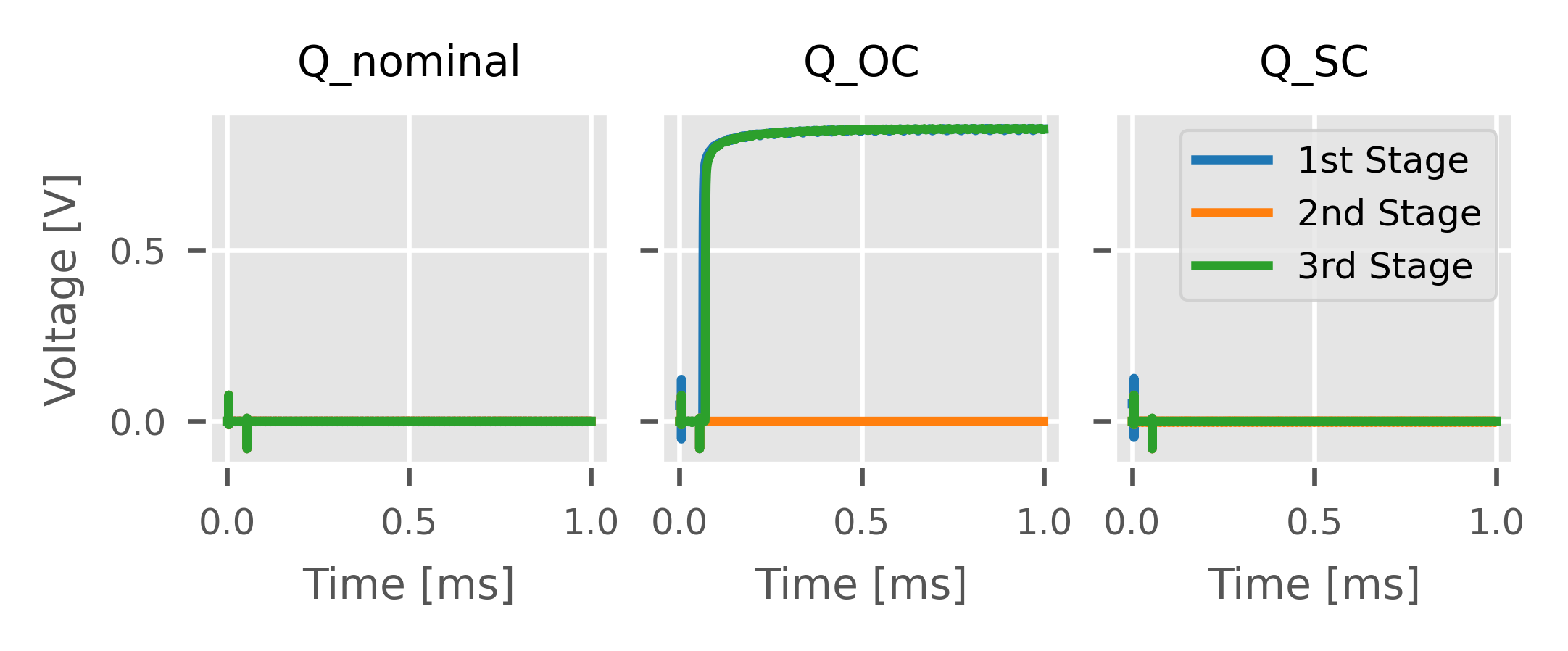}\vspace{-2ex}
    \caption{Output of the BIST after full integration. The first plot corresponds to the nominal operation with no fault detected, the middle plot shows detection of an open-circuit fault, and the last plot illustrates short-circuit fault cases.}
    \label{fig:BIST}
    \vspace{-3ex}
\end{figure}

\begin{figure}
    \centering \includegraphics[width=1\linewidth]{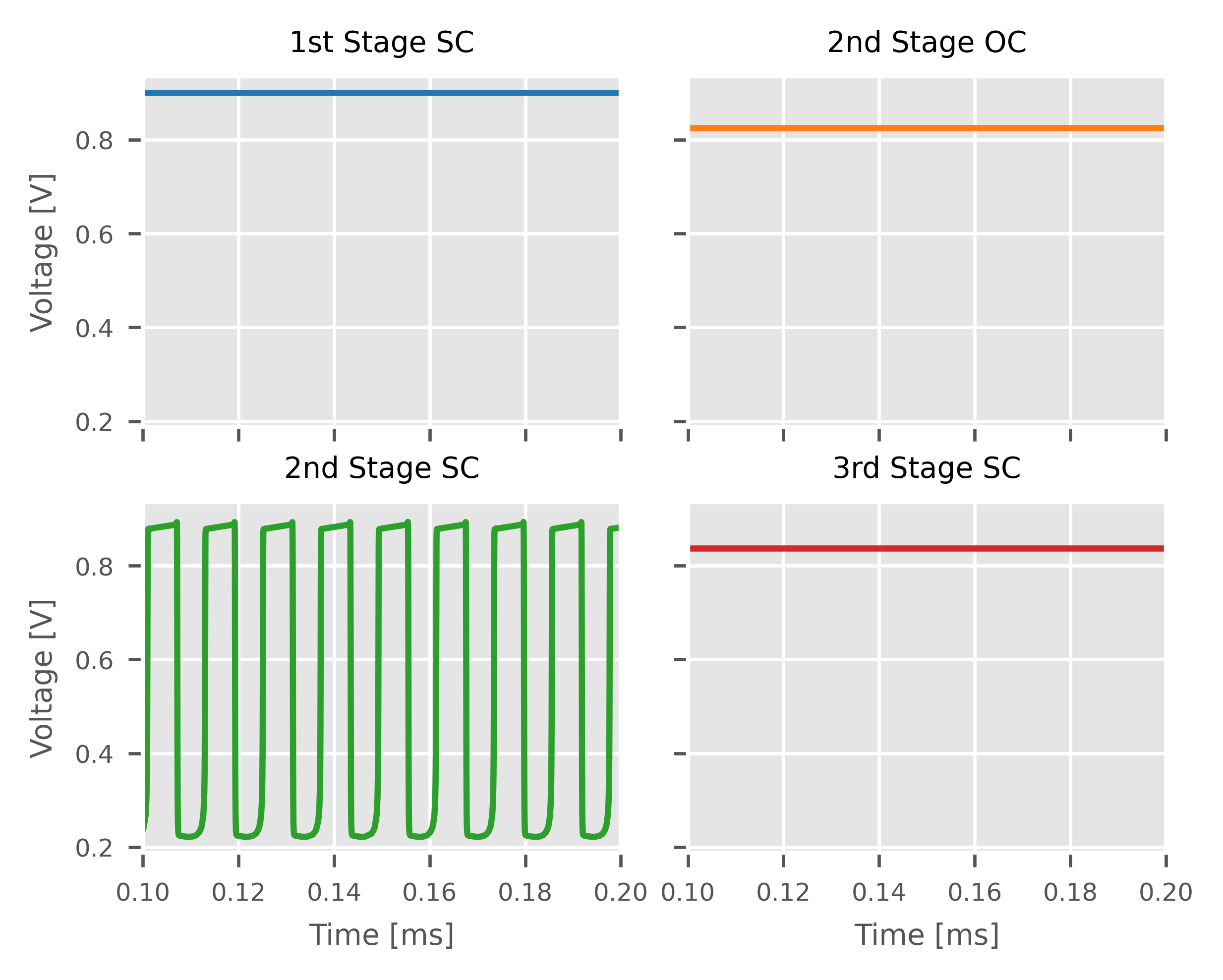}\vspace{-2ex}
    \caption{Detailed VCO outputs highlighting cases where the BIST fault flag was not activated. Shows open-circuit faults and short-circuit faults, explaining fault-tolerant behavior in the second stage.}
    \label{fig:BIST_errors}
    \vspace{-3ex}
\end{figure}

\begin{table}\vspace{-3ex}
\caption{Comparison of ROs and VCOs implemented with FE}
\label{tab:comparison}
\scalebox{1}{\begin{threeparttable}
\resizebox{\linewidth}{!}{%
\centering
\setlength{\arrayrulewidth}{0.4pt} 

\begin{tblr}{
  colspec = {Q[105] Q[35] Q[120] Q[30] Q[70] Q[72]}, 
  vline{2} = {2,3,4,5,6,7,8,9,10}{0.4pt},
  rowsep = 1.15pt, 
  hline{1,2,10} = {1.2pt},
  hline{3,4,5,6,7,8,9} = {-}{0.4pt},
}
\textbf{Ref} & \textbf{V$_{DD}$} & \textbf{Freq.$^1$} & \textbf{Stg$^2$} & \textbf{Power$^3$} & \textbf{Area$^4$} \\

\textbf{IGZO RO (this work$)*$} & \textbf{0.9V} & 40k & \textbf{3} & \textbf{0.0089} & \textbf{0.0041} \\
\textbf{IGZO VCO (this work)$*$} & \textbf{0.9-1.5V} & [69k–98k]-[54k–135k] & \textbf{3} & \textbf{0.0083-0.0165} & \textbf{0.0041} \\
\textbf{IGZO VCO+BIST (this work)$*$} & \textbf{0.9-1.5V} & [70.6k–88.65k]-[55.1k–132.5k] & \textbf{3} & \textbf{0.0091-0.0172} & \textbf{0.0235} \\
IGZO VCO~\cite{VCO_igzo_tejaswini} & 5V & 327–560 & – & 1.3 & Not reported\\
SWCNT VCO~\cite{VCO_Bongjun} & 5V & 0.85k–2.05k & 5 & $\approx$0.4 & $\approx$17.5 \\
IGZO RO~\cite{high-speed_yuanfeng} & 20V & \textbf{1.82M–6.51M} & 7 & $\approx$ 43 & $\approx$ 0.13 \\
IGZO RO~\cite{A-igzo_di} & 25V & 810k & 11 & $\approx$50 & $\approx$0.208 \\
IGZO RO~\cite{fastiIC_arun} & 25V & 406k-2.1M & 5-7 & $\approx$13-43 & $\approx$1.6-4.5 \\

\end{tblr}
}
\begin{tablenotes}\footnotesize
    \item[]$*$At 27ºC and Typical transistor model. $^1$Freq.: frequency (Hz); $^2$Stg: ring stages; $^3$Power: mW; $^4$Area: mm$^{2}$.
\end{tablenotes}
\end{threeparttable}}
\vspace{-3ex}
\end{table}

\textit{Simulation Setup:}
We used Cadence Spectre with device models from PragmatIC's 3rd-generation FlexICs~\cite{EuropracticeFlexibleElectronics}. For the RO, the oscillator operated at 0.9\,V and 27\,$^\circ$C. For the VCO, control voltage was swept from 0.65\,V to 0.85\,V ($V_{\text{DD}}=0.9$\,V) and from 1.05\,V to 1.41\,V ($V_{\text{DD}}=1.5$\,V). The BIST was designed using standard cells from the gen3 FlexIC PDK. Device sizing is presented in Table~\ref{tab:sizing}.


\textbf{RO:} \blue{Evaluated under comprehensive PVT analysis covering process corners (Fast, Typical, Slow), supply voltages ($\pm10\%$ of 0.9\,V), and temperatures (0, 27, 85\,$^\circ$C), remaining functional under all conditions.} Frequency decreased with increasing $V_{\text{DD}}$ (Fig.~\ref{fig:PVT_figure}) because resistive pull-ups do not provide additional current as supply voltage rises, increasing stage delay. The Fast corner achieves up to 122\,kHz (32.8\,$\mu$W); the Slow corner is most power-efficient (5\,$\mu$W); the Typical corner offers a balanced trade-off (Table~\ref{tab:pvt}).

\textbf{VCO:} Achieved frequency ranges of 69–98\,kHz at 0.9\,V and 54.3–135\,kHz at 1.5\,V, with higher oscillation frequencies than the standalone RO without additional circuit overhead.

\textbf{BIST:} \orange{The BIST block occupies 4,585\,$\mu$m$^2$, 12\% larger than the VCO area.} 
Power consumption is 0.251\,$\mu$W at 0.9\,V and 0.426\,$\mu$W at 1.5\,V, adding only 3\% overhead to the RO/VCO power (8–16\,$\mu$W).
\orange{The BIST generates a digital signature based on per-stage delay measurements. Under \blue{defect}-free operation, this signature remains stable across a given PVT corner. A reference signature is established once under nominal conditions (e.g., via a golden device or power-on calibration). While process spread and mechanical bending affect absolute delay values, the \textit{relative} inter-stage delay remains consistent in \blue{defect}-free operation, making the approach robust to manufacturing variations.}
\orange{Post-layout simulations including parasitic extraction confirmed minimal performance degradation (Fig.~\ref{fig:layouts}).}

\textbf{Full Integration (VCO + BIST):} After integrating three BIST blocks (one per stage), the frequency range decreases by approximately 2\% but remains within acceptable limits. Fig.~\ref{fig:BIST} presents the BIST output: the first plot shows nominal operation; the middle plot shows detection of an OC \blue{defect injected at R5 (pull-up of stage~1)}, detected in stages~1 and~3 but not stage~2 (analyzed below); the last plot shows SC \blue{defect} cases where the flag is not activated.
\blue{In the nominal case (first plot), the BIST output correctly remains at logic~0. A flag activation under nominal conditions would constitute a false positive—i.e., a \blue{defect} reported when none is present—confirming that no such spurious detections occur.}

\textbf{\blue{Defect} Detection Evaluation.}
A total of 180 simulations covered all individual \blue{defect} cases and nominal operation. \blue{Defects} were injected at R1–R3 and R5–R7 (SC: 10\,k$\Omega$; OC: 10\,M$\Omega$) across five VCO input voltages at both $V_{\text{DD}}$ values. Combinational \blue{defects} involving two or more simultaneous resistor \blue{defects} were also evaluated.
Fig.~\ref{fig:BIST_errors} shows representative undetected cases. For SC \blue{defects} in stages~1 and~3, the VCO stops oscillating—a \blue{defect} condition by default. For an OC \blue{defect} in stage~2, the output is stuck at logic high. For an SC \blue{defect} in stage~2, the VCO continues oscillating at the same frequency with a different phase, so no flag is raised. \blue{This specific scenario is a primary contributor to the 7\% of individual \blue{defects} that remain undetected: the redundant circuit design and distinct $V_{\text{DD}}$ connection of the middle stage (Fig.~\ref{fig:schematic}b,d) make the circuit inherently tolerant to this \blue{defect}.}

\blue{For multiple simultaneous \blue{defects}, the 88\% coverage (vs.\ 93\% for individual \blue{defects}) reflects cases where two \blue{defects} produce opposing effects on stage delay: one \blue{defect} accelerates a stage while another slows it, causing the XNOR-based comparator to observe a phase shift consistent with fault-free operation. This cancellation effect reduces the net delay deviation below the detection threshold.}

\textit{Comparison with State-of-the-art:}
Table~\ref{tab:comparison} compares the proposed design with reported ROs and VCOs in FE. Our design operates at a lower supply voltage while matching or exceeding prior frequency ranges. \blue{The power comparison reflects the efficiency of the overall RO/VCO architecture; it is included to establish the suitability of the platform for ultra-low-power FE deployment.}
Our solution achieves over 146$\times$ lower power than IGZO-based VCOs~\cite{VCO_igzo_tejaswini} and up to 5600$\times$ lower than other ROs~\cite{A-igzo_di, fastiIC_arun}.

To the best of our knowledge, this is the first fully integrated RO+BIST architecture in IGZO-TFTs, enabling localized in-situ \blue{defect} detection within a functional building block with high \blue{defect} coverage.

\section{Conclusion}\label{sec:conclusion}
Flexible electronics (FE) face challenges such as low carrier mobility, absence of
complementary devices, and high susceptibility to variability-induced \blue{defects},
which complicate reliable analog and mixed-signal circuit design.
This work presents a distributed delay BIST approach that enables modular
\blue{defect} detection \orange{both within the oscillator itself and in upstream
analog/mixed-signal circuits}, achieving high \blue{defect} coverage (93\% for
individual \blue{defects} and 88\% for multiple simultaneous \blue{defects})
\orange{with only 3\% power overhead.}
To support this BIST integration, we developed a compact IGZO-TFT-based RO and
VCO with significant power and area benefits.
\orange{The dual-purpose nature of the VCO+BIST block—serving as both a functional
timing element and a reusable test structure—establishes a scalable strategy for
enhancing reliability in resource-constrained FE systems without requiring external
test equipment.}

{\small
\section*{Acknowledgment}
This work has been supported by the European Research Council (ERC) (Grant No. 101052764) and the KIT International Excellence Fellowship.
}
\bibliographystyle{IEEEtran}
\bibliography{references}

@IEEEtranBSTCTL{IEEEexample:BSTcontrol,
    CTLuse_url = "no",
    CTLuse_doi = "no",
    CTLuse_paper = "yes",
    CTLnames_show_etal = "yes",
    CTLuse_forced_etal = "yes",
    CTLmax_names_forced_etal = "1",
    CTLdash_repeated_names = "no",
}

@article{Zhang2025,
  author = {Zhang, X. and others},
  title = {Design, Fabrication, and Application of Large-Area Flexible Sensor Arrays},
  journal = {MDPI Sensors},
  year = {2025},
  volume = {25},
  number = {6},
  pages = {1234--1245},
  doi = {10.3390/s25061234}
}

@inproceedings{spence:analogfaultsimulation,
author = {Spence, H.},
year = {1996},
month = {10},
pages = {17 - 22},
title = {Automatic analog fault simulation},
isbn = {0-7803-3379-9},
doi = {10.1109/AUTEST.1996.547671}, booktitle={Conference Record. AUTOTESTCON '96}
}

@article{Lim2025,
  author = {Lim, M. S. and Kim, J. and Choi, S.},
  title = {Developments and Future Directions in Stretchable Display Technology},
  journal = {Micromachines},
  year = {2025},
  volume = {16},
  number = {7},
  pages = {772},
  doi = {10.3390/mi16070772}
}

@misc{lu2025testingfaulttolerancetechniques,
      title={Testing and Fault Tolerance Techniques for Carbon Nanotube-Based FPGAs}, 
      author={Siyuan Lu and others},
      year={2025},
      eprint={2508.20304},
      archivePrefix={arXiv},
      primaryClass={cs.AR}, 
}

@inproceedings{LiROtesting,
author = {Li, Katherine Shu-Min and others},
year = {2005},
month = {02},
pages = {184- 187 Vol. 1},
title = {Oscillation ring based interconnect test scheme for SOC},
volume = {1},
isbn = {0-7803-8736-8},
journal = {Proceedings of the Asia and South Pacific Design Automation Conference, ASP-DAC},
}

@inproceedings{bilgic2022performance,
  author    = {Bilgic, Bora and Ozev, Sule},
  title     = {Performance degradation monitoring for analog circuits using lightweight built-in components},
  booktitle = {2022 IEEE 40th VLSI Test Symposium (VTS)},
  pages     = {1-7},
  year      = {2022},
  publisher = {IEEE}
}

@inproceedings{afentaki2025stress,
  author    = {F. Afentaki and others},
  title     = {Exploration of low-power flexible stress monitoring classifiers for conformal wearables},
  booktitle = {Proceedings of the International Symposium on Low Power Electronics and Design (ISLPED)},
  year      = {2025}
}

@article{Jeong2016,
  author    = {J. Woong Jeong and others},
  title     = {Built-in self-test and digital calibration of zero-IF RF transceivers},
  journal   = {IEEE Transactions on VLSI Systems},
  volume    = {24},
  number    = {6},
  year      = {2016}
}

@article{Oshita2016,
  author    = {T. Oshita and others},
  title     = {A compact first-order $\Sigma\Delta$ modulator for analog high-volume testing of complex system-on-chips in a 14nm tri-gate digital CMOS},
  journal   = {IEEE Journal of Solid-State Circuits},
  volume    = {51},
  number    = {2},
  year      = {2016}
}

@INPROCEEDINGS{VCO_igzo_tejaswini,
  author={Keragodu, Tejaswini and others},
  booktitle={2018 IEEE International Symposium on Circuits and Systems (ISCAS)}, 
  title={A Voltage Controlled Oscillator Using IGZO Thin-Film Transistors}, 
  year={2018},
  volume={},
  number={},
  pages={1-5},
  doi={10.1109/ISCAS.2018.8351175}
}

@article{VCO_Bongjun,
author = {Kim, Bongjun and others},
year = {2015},
month = {05},
pages = {},
title = {Voltage-Controlled Ring Oscillators Based on Inkjet Printed Carbon Nanotubes and Zinc Tin Oxide},
volume = {7},
journal = {ACS applied materials \& interfaces},
doi = {10.1021/acsami.5b02093}
}

@ARTICLE{A-igzo_di,
  author={Geng, Di and Kang, Dong Han and Jang, Jin},
  journal={IEEE Electron Device Letters}, 
  title={High-Performance Amorphous Indium–Gallium–Zinc–Oxide Thin-Film Transistor With a Self-Aligned Etch Stopper Patterned by Back-Side UV Exposure}, 
  year={2011},
  volume={32},
  number={6},
  pages={758-760},
  doi={10.1109/LED.2011.2122330}
}

@INPROCEEDINGS{lozano:AKAN,
  author={Lozano Duarte, Paula Carolina and et al.},
  booktitle={2025 26th International Symposium on Quality Electronic Design (ISQED)}, 
  title={Function Approximation Using Analog Building Blocks in Flexible Electronics}, 
  year={2025},
  volume={},
  number={},
  pages={1-7},
}

@ARTICLE{fastiIC_arun,
  author={Suresh, Arun and others},
  journal={IEEE Electron Device Letters}, 
  title={Fast All-Transparent Integrated Circuits Based on Indium Gallium Zinc Oxide Thin-Film Transistors}, 
  year={2010},
  volume={31},
  number={4},
  pages={317-319},
  doi={10.1109/LED.2010.2041525}
}

@ARTICLE{high-speed_yuanfeng,
  author={Chen, Yuanfeng and others},
  journal={IEEE Electron Device Letters}, 
  title={High-Speed Pseudo-CMOS Circuits Using Bulk Accumulation a-IGZO TFTs}, 
  year={2015},
  volume={36},
  number={2},
  pages={153-155},
  doi={10.1109/LED.2014.2379700}
}

@article{Heng:AM2022:FlexHumanMachInterfaces,
author = {Heng, Wenzheng and Solomon, Samuel and Gao, Wei},
title = {Flexible Electronics and Devices as Human–Machine Interfaces for Medical Robotics},
journal = {Advanced Materials},
volume = {34},
number = {16},
pages = {2107902},
year = {2022}
}

@misc{EuropracticeFlexibleElectronics,
  author = {EUROPRACTICE},
  title = {Flexible Electronics},
  year = {2025},
  url = {https://europractice-ic.com/technologies/flexible-electronics/}
}

@inproceedings{Lozano:aspdac25:BinCoDesign,
author={Lozano Duarte, Paula Carolina and others},
booktitle={Asia and South Pacific Design Automation Conference (ASPDAC)},
year={2025},
month={01},
pages={},
title={Design and In-training Optimization of Binary Search ADC for Flexible Classifiers},
doi={10.1145/3658617.3697715}
}

@ARTICLE{defect,
  author={Erozan, Ahmet Turan and others},
  journal={IEEE Trans. on Very Large Scale Integration (VLSI) Systems}, 
  title={Defect Detection in Transparent Printed Electronics Using Learning-Based Optical Inspection}, 
  year={2021},
  volume={29},
  number={8},
  pages={1505-1517}}

\end{document}